\documentclass[12pt]{article}
\usepackage[dvips]{graphicx}
\newcommand{\mathe}{\mathrm{e}}

\begin{document}
\begin{titlepage}
\rightline{\vbox{\halign{&#\cr &GUCAS-CPS-07-01 \cr
&hep-ph/0701264\cr }}} \vskip .7in

\begin{center}

{\Large\bf On the Application of Gluon to Heavy Quarkonium Fragmentation
Functions}

\vskip 0.7in \normalsize {\bf  W.~Qi}$^1$,
{\bf C.-F.~Qiao}$^{2}$, {\bf J.-X.~Wang}$^1$\\
\vskip .5cm

$^1$ Institute of High Energy Physics, the Chinese Academy of Sciences,\\
YuQuan Road 19B, Beijing 100049, China

$^2$ Dept. of Physics, Graduate School of the Chinese
Academy of Sciences,\\
YuQuan Road 19A, Beijing 100049, China \\[2cm]
\end{center}

\begin{abstract}
\normalsize We analyze the uncertainties induced by different
definitions of the momentum fraction $z$ in the application of
gluon to heavy quarkonium fragmentation function. We numerically
calculate the initial $g \rightarrow J / \psi$ fragmentation
functions by using the non-covariant definitions of $z$ with
finite gluon momentum and find that these fragmentation functions
have strong dependence on the gluon momentum $\vec{k}$. As $|
\vec{k} | \rightarrow \infty$, these fragmentation functions
approach to the fragmentation function in the light-cone
definition. Our numerical results show that large uncertainties
remains while the non-covariant definitions of $z$ are employed in
the application of the fragmentation functions. We present for the
first time the polarized gluon to $J/\psi$ fragmentation
functions, which are fitted by the
scheme exploited in this work.\\
\end{abstract}
{\bf \hspace{1cm}PACS number(s):} 13.87.Fh, 14.70.Dj
\end{titlepage}
\section{Introduction}
Fragmentation refers to the process of a parton which carries large
transverse momentum and subsequently forms a jet containing the
expected hadron {\cite{Braaten:1994xb}}. At sufficiently large
transverse momentum of the heavy quarkonium production, the direct
leading order production scheme is normally suppressed while the
fragmentation scheme becomes dominant, though it is formally of
higher order in the strong coupling constant $\alpha_s$
\cite{Braaten:1996pv, Kramer:2001hh, Braaten:1993rw, Lansberg:2006dh}.

Generally, the fragmentation processes of heavy quarkonium $H$
production can be expressed as \cite{Collins:1987}
\begin{eqnarray*}
  d \sigma [A + B \rightarrow H (p_T) + X] & = & \sum_{a, b, c} \int^1_0 d x_a
  \int^1_0 d x_b \int^1_0 d z f_{a / A} (x_a, Q) \times\\
  &  & f_{b / B} (x_b, Q) d \hat{\sigma} (a + b \rightarrow c + X) D_{c
  \rightarrow H} (z, Q),
\end{eqnarray*}
where $a$ and $b$ are incident partons in the colliding hadrons
$A$ and $B$ respectively; $f_{a / A}$ and $f_{b / B}$ are the
parton distribution functions at the scale $Q^2$ of the partonic
subprocess $a + b \rightarrow c + X$; $c$ is the fragmenting
parton (either a gluon or a quark) and the sum runs over all
possible parton contributions; $D_{c \rightarrow H} (z, Q)$ is the
fragmentation function with respect to the scale $Q^2$ which can
be obtained by evolving from the initial fragmentation function
$D_{c \rightarrow H} (z, Q_0)$ using Altarelli-Parisi equations
\cite{Altarelli:1977}
\begin{eqnarray*}
  Q \frac{\partial}{\partial Q} D_{i \rightarrow H} (z, Q) & = & \sum_j
  \int^1_z \frac{d y}{y} P_{ij} (z / y, Q) D_{j \rightarrow H} (y,
  Q)\; ,
\end{eqnarray*}
where $P_{ij}$ are the splitting functions of a parton $i$ into a
parton $j$. The initial gluon fragmentation function $D_{g
\rightarrow H} (z, Q_0)$ is a universal function defined by
factorization in the infinite momentum frame of the fragmenting
gluon. It can also be obtained by calculating a specific physical
process in perturbative QCD in the finite momentum frame of
fragmenting gluon \cite{Braaten:1993rw, Braaten:1995tc}, where the
$z$ is defined in the Lorentz boost invariant form (e.g.
Eq.~(\ref{lightcone}) or (\ref{qiao})). The resulting initial
fragmentation function, independent of the momentum of the parent
gluon, is equivalent to the one derived in the infinite momentum
frame.
\begin{eqnarray}
  z & = & \frac{E^H + p_z^H}{E^g + p^g_z}\; ,  \label{lightcone}\\
  z & = & \frac{p^g \cdot p^H}{(p^g)^2}\; .  \label{qiao}
\end{eqnarray}
In above equations we take the $Z$ axis along the momentum of
fragmenting gluon. Then, $E^H, p_z^H, E^g$ and $p_z^g$ are the
energies and $Z$-components of the four-momenta of the fragmenting
gluon and the produced heavy quarkonium $H$, respectively. In
Eq.~(\ref{lightcone}) $z$ is defined as the usual light-cone form.

The Eqs.~(\ref{lightcone}) and (\ref{qiao}) are hard to be
employed in the application of the gluon fragmentation functions,
because they involve the transverse momentum of the resulting
heavy quarkonium. Instead, usually the non-covariant definitions
as follows are used approximately:
\begin{eqnarray}
  z & = & \frac{\sqrt{M_H^2 + (p_z^H)^2}}{E^g},  \label{appdef1}\\
  z & = & \frac{\sqrt{M_H^2 + (p_z^H)^2} + p_z^H}{E^g + p^g_z},
  \label{appdef2}\\
  z & = & \frac{p_z^H}{p_z^g} .  \label{appdef3}
\end{eqnarray}
When the fragmenting gluon momentum $| \vec{k} | \rightarrow
\infty$, these non-covariant definitions are equivalent to the
light-cone definition in Eq. (\ref{lightcone}). However when the
momentum of the fragmenting gluon remains finite, these
non-covariant definitions of $z$ may induce large uncertainties in
the application of the fragmentation functions. The
Eq.~(\ref{appdef1}) can be re-expressed as
\begin{eqnarray*}
  |p_z^H | & = & \sqrt{(zE^g)^2 - M_H^2}\; ,
\end{eqnarray*}
which shows that at certain values of $z$, there are two
possibilities for momentum directions of quarkonium $H$.
Similarly, the definition of Eq.~(\ref{appdef2}) does not give
unique direction of $H$ for certain $z$, which does not comply
with the original idea of fragmentation. Thus, the definition of
Eq. (\ref{appdef3}) is better and widely used. About a decade ago,
people noticed that the parton to heavy quarkonium fragmentation
functions can be obtained analytically \cite{Braaten:1993rw,
Braaten:1995tc, Ji:1987, Chang:1992ch}. However, in fact the
initial scale gluon fragmentation function can also be obtained
numerically and works equally well as the analytic one in the
application.

In section 2, by using the perturbative QCD we numerically
calculate the initial $g \rightarrow J / \psi$ fragmentation
functions $D_{g \rightarrow J / \psi} (z, M_{J / \psi})$ with the
non-covariant definitions given above at finite momentum of
fragmenting gluon. We compare them with the fragmentation
functions in the light-cone definition Eq.~(\ref{lightcone}). In
section 3, We give out numerically the polarized gluon to $J/\psi$
fragmentation functions with light-cone definition of $z$, and
transform it into analytic ones by fitting. We also present the
$\chi$-square fitted functions for both polarized and unpolarized
fragmentation functions in the light-cone definition of $z$. In
section 4, we give a brief summary of our results.

\section{Unpolarized fragmentation functions w.r.t. different
definitions of $z$ and the resulting uncertainties}

\begin{figure}
\centering
\includegraphics[scale=0.4]{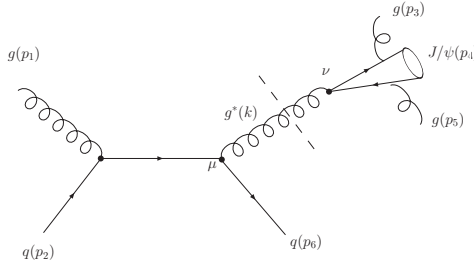}
  \caption{The Feynman diagram of process $qg \rightarrow qg^{\ast}
  \rightarrow J / \psi gg$\label{figfragproc}}
\end{figure}

The Feynman diagram of the process $qg \rightarrow qg^{\ast}
\rightarrow J / \psi gg$ is shown in Fig.~\ref{figfragproc} and
the corresponding matrix element is composed of two parts: (1) the
production of a virtual gluon $g^{\ast}$ with momentum $k$ and
invariant mass $\sqrt{s}$; (2) the virtual gluon decaying into a
$J / \psi$ and a gluon pair. i.e.,
\begin{eqnarray}
\mathcal{M} & = & M_1^{\mu} \left( \frac{g_{\mu
\nu}}{k^2} \right) M_2^{\nu} (\kappa),\nonumber\\
& = & - \sum_{\lambda} M_1^{\mu} \varepsilon_{\mu} (\lambda)
\frac{1}{k^2}
M_2^{\nu}(\kappa) \varepsilon_{\nu}^{\ast} (\lambda),
\end{eqnarray}
where $M_1^{\mu} \varepsilon_{\mu} (\lambda)$ and $M_2^{\nu} (s)
\varepsilon^{\ast}_{\nu} (\lambda)$ represent the virtual gluon
production and decay sectors, respectively. The $\lambda$ and
$\kappa$ denote the virtual gluon and $J / \psi$ polarization
indices. In deriving out the above second equation, the following
condition on the summation of gluon polarization vector is used:
\begin{eqnarray}
  g_{\mu \nu}^{} & = & - \sum_{\lambda} \varepsilon_{\mu} (\lambda)
  \varepsilon^{\ast}_{\nu} (\lambda)\; .
\end{eqnarray}
Then, the differential cross-section of process $g + q \rightarrow q
+ g^*; g^* \rightarrow J/\psi + g + g $, as shown in
Fig.~\ref{figfragproc}, can be expressed as:
\begin{eqnarray}
  d \sigma & = & |\mathcal{M}|^2 [d \phi] \nonumber\\
  & = & \int d^3 k \frac{d \hat{\sigma} (qg \rightarrow qg^{\ast})}{d^3 k}
  \int_{M_{J / \psi}^2}^{\infty} \frac{ds}{\pi s^{3 / 2}} \sum _\kappa d
  \Gamma_\kappa (g^{\ast} \rightarrow J / \psi gg),
\end{eqnarray}
where the sum runs over the $J/\psi$ polarization for the
unpolarized production, and
\begin{eqnarray}
  d \hat{\sigma} (qg \rightarrow qg^{\ast}) & = & \delta^4 (p_1 + p_2 - k -
  p_6) \sum_{\lambda} |M_1 |^2_{\lambda \lambda} [d \phi_1] \frac{d^3 k}{(2
  \pi)^3 2 E^g},\\
  d \Gamma_\kappa (g^{\ast} \rightarrow J / \psi gg) & = & \frac{(2 \pi)^4}{2
  \sqrt{s}} \delta^4 (k - p_3 - p_4 - p_5) \left( \frac{1}{2} \sum_{\lambda}
  |M_2 (\kappa) |^2_{\lambda \lambda} [d \phi_2] \right),\\
  \left[ d \phi \right]  =  \left[ d \phi_1 \right] \left[ d \phi_2
  \right]\!\!\!\!\!&,&
  \left[ d \phi_1 \right]  =  \frac{d^3 p_6}{(2 \pi)^3 2 E_6}\; ,
  \;\;\;
  \left[ d \phi_2 \right]  =  \prod_{i = 3, 4, 5} \frac{d^3 p_i}{(2 \pi)^3 2
  E_i} \; .
\end{eqnarray}
Hence, the initial scale gluon fragmentation function corresponding
to the finite gluon momentum satisfies
\begin{eqnarray}
  \int_0^1 dz D_{g \rightarrow J / \psi} (z, M_{J / \psi}) & = & \int dp_{4 z}
  \int_{M_{J / \psi}^2}^{\infty} \frac{ds}{\pi s^{3 / 2}} \sum_\kappa
  \int \frac{d \Gamma_\kappa (g^{\ast} \rightarrow J / \psi gg)}{dp_{4 z}}\; .
  \label{fragunpolar}
\end{eqnarray}
Therefore, from Eq.~(\ref{fragunpolar}) one can extract the
unpolarized initial gluon fragmentation functions with either
boost-invariant or non-invariant definitions of $z$. We numerically
calculate the fragmentation functions with various definitions of
$z$ with a specific gluon momentum and compare the differences among
them.

\begin{figure}
\centering
\includegraphics[scale=0.4]{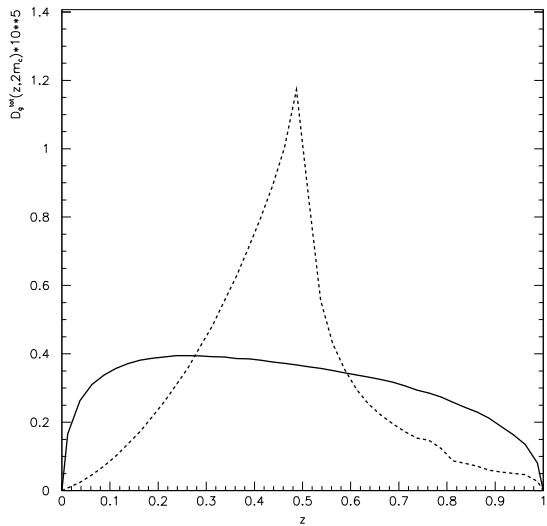}
\caption{The fragmentation functions w.r.t. Eq.~(\ref{lightcone})
(solid line) and (\ref{qiao}) (dash line)} \label{ligqiao}
\end{figure}

Fig. \ref{ligqiao} is the fragmentation functions corresponding to
the boost-invariant definitions of $z$ given in
Eq.~(\ref{lightcone}) and (\ref{qiao}), which are numerically
calculated using the FDC (Feynman Diagram Calculation) program
{\cite{Wang:2004du}} and they agree with those given in Refs.
{\cite{Braaten:1993rw}} and {\cite{Qiao:1997wb}}, respectively.
Given different virtual gluon momentums in the numerical
calculation, we obtain the same fragmentation function
distribution versus $z$ as shown in Fig.~\ref{ligqiao}, which
indicates explicitly the independence of the fragmentation
functions on virtual gluon momentum, as they should be from the
definitions of $z$.

\begin{figure}
\centering
\includegraphics[scale=0.4]{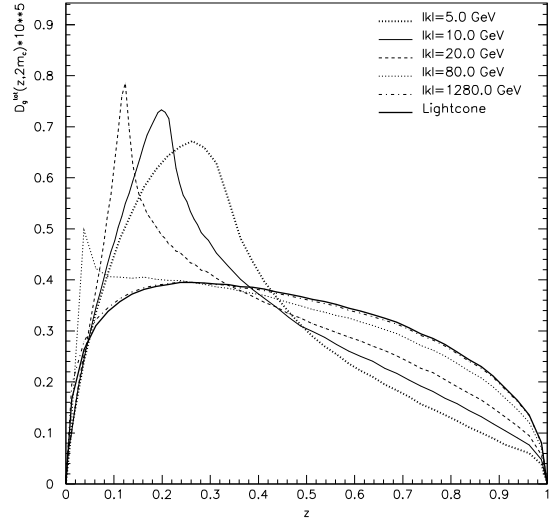}
  \caption{Fragmentation functions w.r.t. Eq.~(\ref{appdef1})
  \label{figappdef1}}
\end{figure}

\begin{figure}
\centering
\includegraphics[scale=0.4]{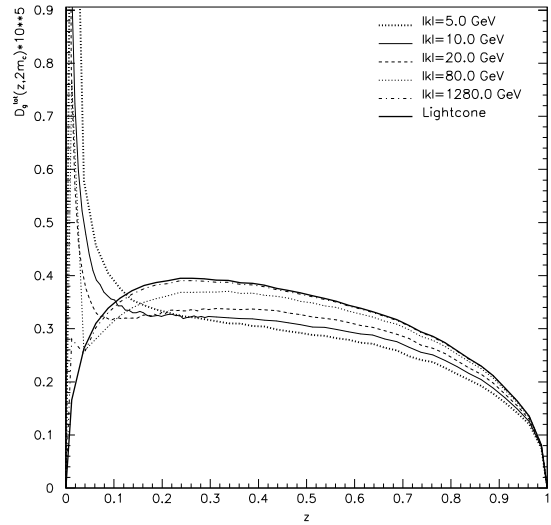}
\caption{Fragmentation functions w.r.t. Eq.~(\ref{appdef2})
\label{figappdef2}}
\end{figure}

Figs. \ref{figappdef1} and \ref{figappdef2} present the
fragmentation function distributions over variable $z$ in
non-covariant definition of Eqs.~(\ref{appdef1}) and
(\ref{appdef2}). It shows that the fragmentation function
distributions rely on the momentum $| \vec{k} |$ of the virtual
gluon. When $| \vec{k} | \rightarrow \infty$, the corresponding
fragmentation functions approach to the one in the light-cone
definition of Eq.~(\ref{lightcone}). However, when $| \vec{k} |$ is
finite, these fragmentation functions distribute distinctively from
the one in the light-cone definition, which will bring certain
uncertainties in the application of the fragmentation functions.
With the non-covariant definition in Eqs. (\ref{appdef1}) and
(\ref{appdef2}), there is the possibility that the $J / \psi$
momentum is opposite to the virtual gluon momentum $\vec{k}$, and
this possibility becomes larger as $|\vec{k}|$
decreases. Therefore, the large peaks at small $z$ in
Figs.~(\ref{figappdef1}) and (\ref{figappdef2}) correspond to the
lower momentum $| \vec{k} |$.

\begin{figure}
\centering
\includegraphics[scale=0.4]{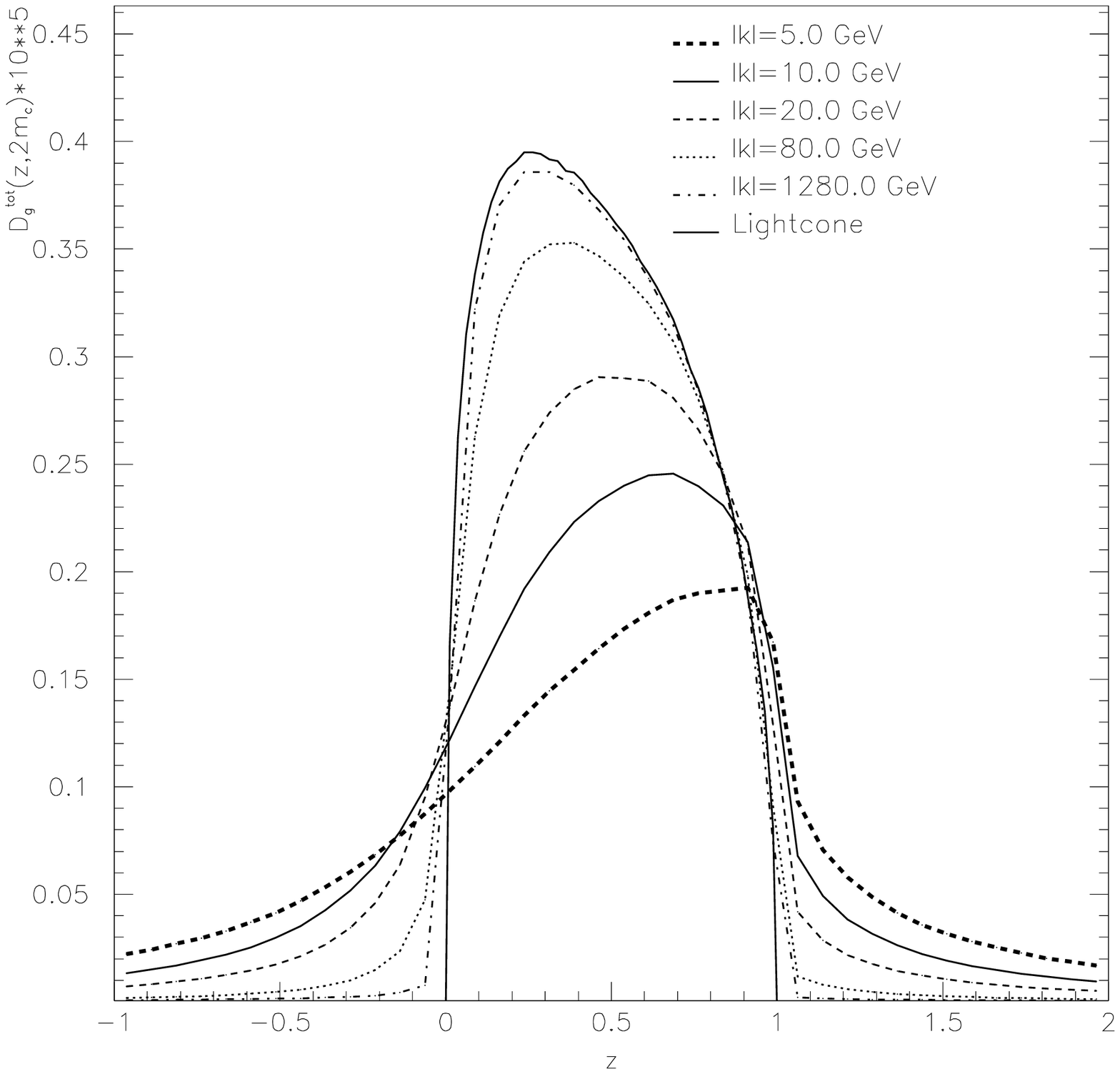}
  \caption{Fragmentation function w.r.t. Eq.~(\ref{appdef3})
  \label{figappdef3}}
\end{figure}

Figure \ref{figappdef3} gives the distribution of the
fragmentation functions versus $z$ in the definition of
(\ref{appdef3}) at various virtual gluon momentums. Since it is
possible that the $J / \psi$ may possess momentum opposite to that
of the virtual gluon, in the definition of (\ref{appdef3}), the
condition of $z < 0$ exists. And because of the non-zero invariant
mass of the virtual gluon, even $z
> 1$ happens. Thus, in Fig.~\ref{figappdef3} the fragmentation functions
distribute in the scope of $- 1 \leq z \leq 2$. As shown in the
figure, when the virtual gluon momentum $| \vec{k} | \rightarrow
\infty$, fragmentation functions approach to the one in the
light-cone definition. When $| \vec{k} | = 5, 10, 20, 80$ and
$1280$GeV, the ratios of fragmentation probabilities in the
definition of (5), that is the integrations of the corresponding
fragmentation functions over $z$, to the one in light-cone
definition are 0.49, 0.65, 0.77, 0.91 and 0.97 respectively in the
range of $0 \leq z \leq 1$. This indicates that the smaller the
virtual gluon momentum is, the more possible the produced $J /
\psi$ will be opposite to the gluon. Therefore, uncertainties
occur in the application of the fragmentation function in the
definition of (\ref{appdef3}) when gluon momentum $|\vec{k}|$ is
not big enough. While the virtual gluon energy is greater than
80GeV, the uncertainties induced by the definition of
(\ref{appdef3}) will drop to less than 10\%.

\begin{figure}
\centering
\includegraphics[scale=0.4]{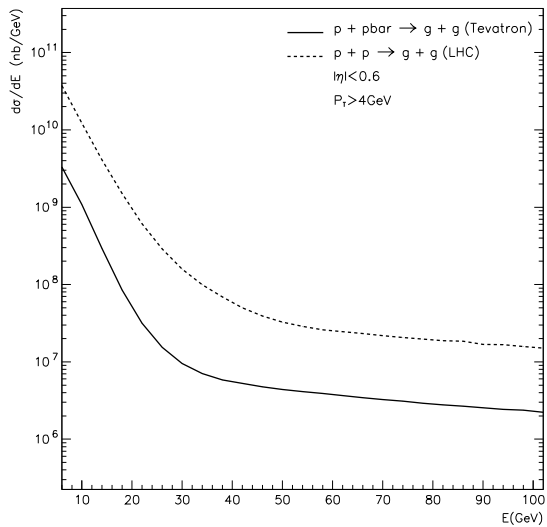}
\caption{The energy dependence of the differential cross-sections
of $p\; p(\bar{p})\rightarrow g\; g$ process at LHC and Tevatron.
The horizontal variable E denotes the final gluon energy.}
\label{figcrosec}
\end{figure}

Figure \ref{figcrosec} presents the differential cross-sections of
processes $p\; p(\bar{p}) \rightarrow g\; g$ versus gluon energy at
the colliding energies of Fermilab Tevatron and LHC respectively.
In our numerical calculation, the parton distribution function of
CTEQ6L \cite{cteq6} is employed. As shown in the figure, the
cross-section decreases quite fast as the gluon momentum
increases. Whereas, from the analysis above, we know that the at
some low energies of the fragmenting gluon there are large
discrepancies between the fragmentation functions with
non-covariant definitions and boost-invariant definitions of $z$.
Therefore, it is possible that the non-covariant definitions of
$z$ may induce large errors in the application of fragmentation
functions.

\section{Polarized fragmentation functions w.r.t. light-cone definition of $z$}
The polarized fragmentation functions of gluon fragmenting into
P-wave charmonium $\chi_{cJ}$ exist in the literature
\cite{Cho:1995mb}, however, there has been no relevant work on
polarized fragmentation functions of gluon to S-wave vector
Quarkonium states. With the FDC program and procedure described in
above, in fact it is straightforward for us to obtain the gluon to
polarized $J/\psi$ fragmentation functions numerically. And then
get the analytic ones by fitting, which are equivalent in use as
the ones obtained directly from the analytic calculation, at least
to large extent.

\begin{figure}
\centering
\includegraphics[scale=0.4]{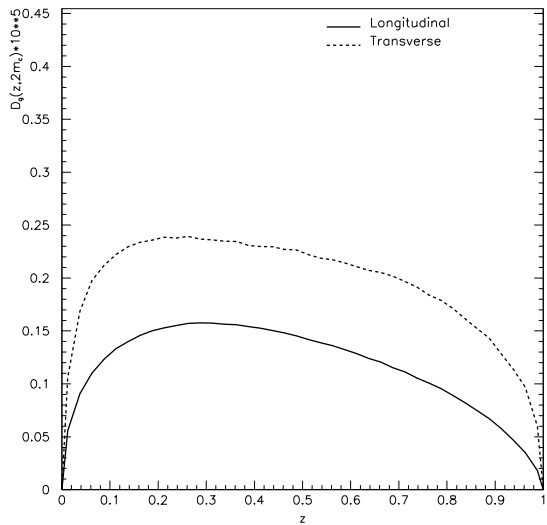}
  \caption{Polarized fragmentation functions w.r.t. to Eq.
  \ref{appdef1}\label{polarfrag}}
\end{figure}

With the numerical method described in Sec.~2, we can also obtain
the polarized fragmentation functions which satisfy
\begin{eqnarray}
  \int_0^1 dz D_{g \rightarrow J / \psi}^T (z, M_{J / \psi}) & = & \int dp_{4 z}
  \int^{\infty}_{M_{J / \psi}^2} \frac{ds}{\pi s^{3 / 2}} \sum_{\kappa = 1, 2} \int
  \frac{d \Gamma_\kappa (g^{\ast} \rightarrow J / \psi gg)}{dp_{4 z}},
  \label{tranpolar}\\
  \int_0^1 dz D_{g \rightarrow J / \psi}^L (z, M_{J / \psi}) & = & \int dp_{4 z}
  \int^{\infty}_{M_{J / \psi}^2} \frac{ds}{\pi s^{3 / 2}} \int \frac{d
  \Gamma_3 (g^{\ast} \rightarrow J / \psi gg)}{dp_{4 z}},  \label{longpolar}
\end{eqnarray}
where $D^T_{g \rightarrow J / \psi}$ and $D^L_{g \rightarrow J /
\psi}$ are fragmentation functions of gluon to $J/\psi$ in
transverse and longitudinal polarizations, respectively. The
polarized fragmentation functions can then be extracted from
Eqs.~(\ref{tranpolar}) and (\ref{longpolar}) in different
definitions of $z$. Fig. \ref{polarfrag} shows the polarized
fragmentation functions corresponding to the light-cone definition
of Eq.~(\ref{lightcone}). Explicit numerical calculation indicates
that these polarized fragmentation functions in light-cone
definition are independent of the virtual gluon momentum, as they
should be.

In Ref.~\cite{Braaten:1995tc}, the initial unpolarized
fragmentation function of gluon to $J/\psi$ is given in a form of
two-dimensional integrals, which in practice can only be evaluated
numerically and thus is not very convenient for the application of
the Altarelli-Parisi evolution equation \cite{Altarelli:1977}. In
use of the obtained initial gluon to unpolarized $J/\psi$
fragmentation function \cite{Braaten:1995tc}, one can integrate out
the integrals numerically \cite{Qi:2006} and fit the $z$ distribution into a
combination of functions of $z$. Then the application of the gluon
to $J/\psi$ fragmentation function will be more easier for the
phenomenological use. To our knowledge the polarized fragmentation
functions of gluon to $J/\psi$ are still absent so far in 
literatures. However, they may be necessary for future careful
analysis about the polarization situation of Quarkonium
production.

We proceed the $\chi$-square fitting of gluon to polarized and
unpolarized $J/\psi$ fragmentation functions into polynomials and
exponents in the light-cone definition of $z$, which tells
\begin{eqnarray}
D_{g \rightarrow J / \psi} (z, M_{J / \psi}) & = &
\alpha_s^3(M_{J/\psi})\frac{|R(0)|^2}{(\frac{1}{2}M_{J/\psi})^3}\mathe^{a_0
z (1 - z)} (1
- z) \sum_{n = 1, 9} a_n z^n\; , \\
D^T_{g \rightarrow J / \psi} (z, M_{J / \psi}) & = &
\alpha_s^3(M_{J/\psi})\frac{|R(0)|^2}{(\frac{1}{2}M_{J/\psi})^3}\mathe^{a^T_0
z (1 -
  z)} (1 - z) \sum_{n = 1, 9} a^T_n z^n\; .
\end{eqnarray}
Here, $\alpha_s(M_{J/\psi})$,  $M_{J/\psi}$ and $R(0)$ are strong
coupling, the $J/\psi$ mass and radial wave function at the origin,
respectively. The fitted coefficients of $a_n$ and $a_n^T$ are
presented in Table \ref{coeftbl}.
\vskip 0.5cm
\begin{table}[h]
\centering
  \begin{tabular}{|c|c||c|c|}
    \hline
    $a_0$ & $ -17.0402262                    $ & $a_0^T$ & $ -17.3527537                 $  \\
    $a_1$ & $ 5.1820321     \times 10^{-7}   $ & $a_1^T$ & $ 3.3809309    \times 10^{-7} $  \\
    $a_2$ & $ -7.5744212    \times 10^{-6}   $ & $a_2^T$ & $ -5.0184819   \times 10^{-6} $ \\
    $a_3$ & $ 1.2310370     \times 10^{-4}   $ & $a_3^T$ & $ 8.2006550    \times 10^{-5} $ \\
    $a_4$ & $ -7.6536059    \times 10^{-4}   $ & $a_4^T$ & $ -5.1990241   \times 10^{-4} $ \\
    $a_5$ & $ 3.1398595     \times 10^{-3}   $ & $a_5^T$ & $ 2.1421231    \times 10^{-3} $ \\
    $a_6$ & $ -7.3523074    \times 10^{-3}   $ & $a_6^T$ & $ -4.9941798   \times 10^{-3} $ \\
    $a_7$ & $ 9.2419267     \times 10^{-3}   $ & $a_7^T$ & $ 6.2434863    \times 10^{-3} $ \\
    $a_8$ & $ -5.8527342    \times 10^{-3}   $ & $a_8^T$ & $ -3.9354343   \times10^{-3}  $ \\
    $a_9$ & $ 1.4728577     \times 10^{-3}   $ & $a_9^T$ & $ 9.8679226    \times 10^{-4} $ \\
    \hline
\end{tabular}
\caption{Coefficients of fitted functions for the unpolarized and
transversely polarized initial gluon to $J/\psi$ fragmentation
functions} \label{coeftbl}
\end{table}

\section{Summary}
In this work, we show that although the fragmentation functions
corresponding to the boost invariant definitions of
(\ref{lightcone}) and (\ref{qiao}) are equivalent to the ones in
the infinite momentum frame of gluon and thus independent of
fragmenting gluon momentum, they are inconvenient in use, because
they involve the transverse momentum of the parent gluon. Instead,
the non-covariant definitions, such as in Eqs.~(\ref{appdef1}),
(\ref{appdef2}) and (\ref{appdef3}), are used as an approximation.
They are equivalent to the definition of the light-cone coordinate
form when the fragmenting gluon momentum $| \vec{k} | \rightarrow
\infty$. However, as shown in our calculation, these non-covariant
definitions may induce large errors while $| \vec{k} |$ is finite,
especially small. In practice, the definition of (\ref{appdef1})
and (\ref{appdef2}) are somewhat inconvenient because they
possess, in the range of $0 \leq z \leq 1$, the possibility of $J
/ \psi$ moving in the opposite direction of the fragmenting gluon.
The definition of (\ref{appdef3}) excludes this situation, but it
may bring larger errors in the colliding energies of Tevatron and
LHC. This should be taken into account in the careful
phenomenological analysis in the application of the gluon
fragmentation functions.

In this paper we give out the initial fragmentation functions of
gluon into the polarized $J/\psi$ for the first time. We use a
different method in getting them from the normal analytic
calculation in the literature. The $\chi$-square fittings for both
polarized and unpolarized fragmentation functions of gluon into
$J/\psi$ are done and presented, which are suitable for future
phenomenological use. Finally, although the polarized
fragmentation functions obtained in this work are schematically
for charmonium, the $J/\psi$, in fact they can be directly applied
to the $\Upsilon$ system with some simple replacements. 
\vskip 0.7cm

\centerline{\bf \large Acknowledgments}

The work of  C.~F.~Q. was supported in part by the Natural Science
Foundation of China and by the Scientific Research Fund of GUCAS
(No. 055101BM03); the work of J.~X.~W. and W.~Q. was supported by the
Natural Science Foundation of China (No. 10475083).

\end{document}